\documentclass[prb,amsmath,amssymb,twocolumn,showpacs,floatfix]{revtex4}
\usepackage{float}
\usepackage[caption = false]{subfig}
\usepackage{graphicx}
\usepackage{bm}

\usepackage{color}

\begin{document}

\title{Charge segregation model for superconducting correlations in cuprates above  $T_c$}

\author{E.V.L. de Mello}
\affiliation{Instituto de F\'{\i}sica, Universidade Federal Fluminense, Niter\'oi, RJ 24210-340, Brazil}
\author{J.E. Sonier}
\affiliation{Department of Physics, Simon Fraser University, Burnaby, British Columbia, Canada V5A 1S6}

\date{\today}

\begin{abstract}
We present a theoretical framework for understanding recent transverse field muon spin rotation (TF-$\mu$SR) 
experiments on cuprate superconductors in terms of localized regions of phase-coherent pairing correlations 
above the bulk superconducting transition temperature $T_c$. The local regions of phase coherence are associated 
with a tendency toward charge ordering, a phenomenon found recently in hole-doped
cuprates. We use the 
Cahn-Hilliard equation as a means to phenomenologically model the inhomogeneous 
charge distribution of the electron system observed
experimentally. For this system we perform self-consistent 
superconducting calculations using the Bogoliubov-deGennes method.  
Within this context we explore two possible scenarios: (i) The magnetic field is diamagnetically screened by the sum of 
varying shielding currents of isolated 
small-sized  superconducting 
domains. (ii) These domains become 
increasingly correlated by Josephson coupling as the temperature is lowered 
and the main response to the applied magnetic field is from the sum of
all varying tunneling currents.
The results indicate that these two approaches may be used to simulate
the TF-$\mu$SR data but case (ii) yields better agreement.

\end{abstract}

\maketitle

\section{Introduction}

Determining the relationship between the pseudogap phase and the superconducting (SC) properties of hole-doped cuprates 
has been a primary challenge to understanding high-$T_c$ superconductivity. Some time ago, Emery and Kivelson 
suggested that the phase order associated with the complex order parameter of the SC state goes away at
a temperature proportional to the superfluid density, and vanishes  below the lower critical doping for bulk 
superconductivity.\cite{EmeryNat} The pseudogap phase in their model is associated with preformed Cooper 
pairs, but superconductivity is inhibited by large phase fluctuations. To date there are a number of
experiments \cite{Corson,Ong1,Ong,Bilbro,Grbic,Kanigel,Gomes} that provide compelling evidence for the existence of SC
fluctuations within the pseudogap state of hole-doped cuprates above $T_c$.

 While in the past it was often assumed that such SC fluctuations reside in a 
homogeneous environment, recent X-ray scattering and
scanning tunneling microscopy (STM) experiments have collectively established that the normal state of underdoped cuprates also
includes incommensurate charge order (CO) fluctuations.\cite{Wise,Ghiringhelli,Chang,Torchinsky,LeTacon,Comin,Neto} 
X-ray scattering measurements
show the incommensurate CO emerging just below the pseudogap temperature $T^*$ in Bi$_2$Sr$_{2-x}$La$_x$CuO$_{6+\delta}$,\cite{Comin} but
far below $T^*$ in underdoped YBa$_2$Cu$_3$O$_{6+x}$ (YBCO).\cite{Ghiringhelli,Chang}
The CO fluctuations are two-dimensional in that they are characterized by a substantially longer maximum correlation
length in the in-plane direction ({\it i.e.} 20 to 100~\AA), and in X-ray 
experiments have been demonstrated to compete with superconductivity below
$T_c$.\cite{Ghiringhelli}  These experimental findings indicate a universal intrinsic incipient charge instability in high-$T_c$ cuprate superconductors,
which may play a pivotal role in limiting the temperature extent of bulk superconductivity.

Recently, transverse-field muon spin rotation (TF-$\mu$SR) 
experiments have  also provided evidence for an 
intrinsic source of electronic inhomogeneity in
superconducting cuprates above $T_c$.\cite{Sonier:08,Mahyari} A generic spatially inhomogeneous response to an applied
magnetic field is detected in La$_{2-x}$Sr$_x$CuO$_4$ (LSCO), 
Bi$_2$Sr$_2$CaCu$_2$O$_{8+\delta}$ (BSCCO) and YBCO, extending to temperatures 
well above $T_c$. 
The experimental observation is a residual depolarization of the TF-$\mu$SR time spectrum, quantified  
by an exponential relaxation rate $\Lambda$, which is proportional to the half width at half maximum (HWHM)
of a Lorentzian distribution of internal magnetic field. The proportionality constant is
the muon gyromagnetic ratio $\gamma_{\mu}$. The quantity $\Lambda$ increases for stronger
applied field, tracks $T_c$ as a function of hole doping $p$, and in YBCO is reduced near $p \! = \! 1/8$. Moreover, 
$\Lambda$ is found to scale with the maximum value of $T_c$ for each family of 
compounds. These observations
are explained by the occurrence of inhomogeneous SC fluctuations above $T_c$, driven by
an intrinsic rearrangement of the electronic structure\cite{Jeff} --- presumably associated with 
a competing order of some kind and likely related to 
the CO instability mentioned above. 

Hayward et al.\cite{Hayward1} have recently modeled competing CO and SC correlations in the pseudogap
phase by angular fluctuations of a multi-component order parameter. The model reproduces the
temperature dependence of the X-ray scattering intensity of CO correlations in YBa$_2$Cu$_3$O$_{6.67}$,
and has been used to characterize the contribution of SC fluctuations to the bulk diamagnetic
susceptibility.\cite{Hayward2} However, this model does 
not describe the spatial inhomogeneity that may arise from these
competing orders.\cite{Hayward2}
Here we propose a model of spatially static segregated CO and SC
correlations in the pseudogap phase to describe the hole-doping dependence of the TF-SR data,
with an understanding that in the absence of fluctuations our theory will not 
accurately reproduce the
temperature dependence of the TF-$\mu$SR relaxation rate.

To address early evidence of spatial charge inhomogeneity in cuprate superconductors, \cite{Muller} one of us proposed that there exists an electronic phase separation transition 
associated with two competing phases having the same Ginzburg-Landau free energy minima.\cite{PhysC12,EPL12a,EPL12b} 
Such a transition can be studied by the time dependent Cahn-Hilliard (CH) equation, since the 
average doping level is independent of temperature. The two phases develop as small-sized regions 
of low and high charge density, which may favor the formation of localized Cooper pairs --- a situation 
akin to that of a granular superconductor.\cite{Ketterson}. 
This is possible because the SC coherence length of cuprate superconductors is quite short, and 
of the order of the CO correlation length. 

Here we apply the same approach to provide a theoretical framework for understanding the behavior
of $\Lambda$ from the TF-$\mu$SR experiments. The paper is organized as follows:
In Sec.~II we describe simulations of charge inhomogeneity  in small regions 
or patches using the time-dependent CH equation.
Then, in Sec.~III we present the results of calculations of the local $d$-wave
pairing amplitude
by the Bogoliubov-deGennes method. The main purpose of these two sections
is to provide a theoretical framework for simulating the distribution of local
pairing gaps observed by STM  on BSCCO. In Sec.~IV we apply the critical-state Bean model with a spatially varying
shielding current density $J_c({\bf r})$ that is related to the local 
$d$-wave pair potential $\Delta_d({\bf r})$ to describe the spatial dependence of the local magnetic field
$B({\bf r})$ within the superconducting regions or patches. 
The calculated  line width of the corresponding
field distribution is then shown to differ somewhat from the hole-doping dependence
of the TF-$\mu$SR line width observed above $T_c$ in Ref.~\onlinecite{Mahyari}. 
In Sec.~V we consider another approach; the small-sized superconducting 
domains form a granular superconductor coupled by
the Josephson coupling energy $E_j$, which increases as the temperature is lowered 
towards $T_c$. We show that this approach is
able to provide a better description of the doping evolution of the
TF-$\mu$SR data above $T_c$. 

\section{Simulation of Charge Inhomogeneity}

 In this section we use the CH approach 
to  phenomenologically simulate an inhomogeneous charge distribution that
is compatible with experimental findings. 

Even before the recent detection of incommensurate CO
there were experimental indications that nanoscale phase separation is a 
universal phenomenon of cuprates,\cite{Muller} as well as other strongly correlated 
electron systems like manganites.\cite{Dagotto} Motivated by these observations, 
one of us developed a theory in which
the free energy of a system 
with average charge-carrier doping $p$ may be lowered by charge segregation into low 
and high doping regions.\cite{PhysC12,Mello09} The temperature
$T_{\rm PS}(p)$ at which such phase separation occurs was evaluated in these prior calculations,
and found to increase with decreasing $p$ in a manner closely resembling the pseudogap
temperature $T^*(p)$.\cite{TS} This implies that the degree of charge disorder 
diminishes with increased doping --- a result which is implicit in our simulations here, and
a trend which is consistent with a variety of experiments.\cite{Bozin,Singer,McElroy} 

A common way to study the charge segregation process is by way of the Ginzburg-Landau (GL) free 
energy density functional and its usual order parameter power expansion. The convenient order 
parameter for this approach is the normalized difference between the local and the average hole doping, $u({\bf r}_i,t)=[p({\bf r}_i,t)-p]/p$. 
This order parameter is incompatible with  a doping-dependent CO correlation length. However, since the 
latter appears to change very weakly with 
doping,\cite{Comin,Hucker} this approach captures the main effect of a nanoscale CO instability.
The GL free energy density in this
formalism is
\begin{equation}
f(u,T)= {{\frac{1}{2}\varepsilon^2 |\nabla u|^2 +V_{GL}(u,T)}},
\label{FE}
\end{equation}
where the potential is defined as ${\it V_{GL}}(u,T)= -A^2(T)u^2/2+B^2u^4/4+...$,
$A^2(T)=\alpha[T_{\rm PS}(p)-T]$, $\alpha$ and $B$ are constants, and 
$\varepsilon$ controls  the size of the interface between neighboring 
patches.\cite{Otton,Mello04} 

To describe the growth and development of  spatial charge 
inhomogeneity in the CuO$_2$ 
planes, we use the time-dependent CH approach to determine the time evolution of the 
phase separation transition.\cite{CH} 
The CH equation can be written in the form of the following continuity equation 
for the local free energy current density  ${\bf J}=M{\bf\nabla}(\delta f/ \delta u)$,\cite{Bray}

\begin{eqnarray}
\frac{\partial u}{\partial t} & = & -{\bf \nabla} \cdot {\bf J} \nonumber \\
& = & -M\nabla^2[\varepsilon^2\nabla^2u
- A^2(T)u+B^2u^3],
\label{CH}
\end{eqnarray}
where $M$ is the mobility or the charge transport coefficient that sets both the phase separation time scale and
the contrast between the values of $u$ for the two phases. The order parameter varies
between $u({\bf r}_i,t)\approx 0$ corresponding to the homogeneous system above $T_{\rm PS}$, and
$u({\bf r}_i,t\rightarrow \infty) =  \pm  \:1$ corresponding to the extreme case 
(near $T\approx 0$ K) of complete phase separation.

\begin{figure}[ht]
\begin{center}
     \centerline{\includegraphics[width=6.50cm,angle=0]{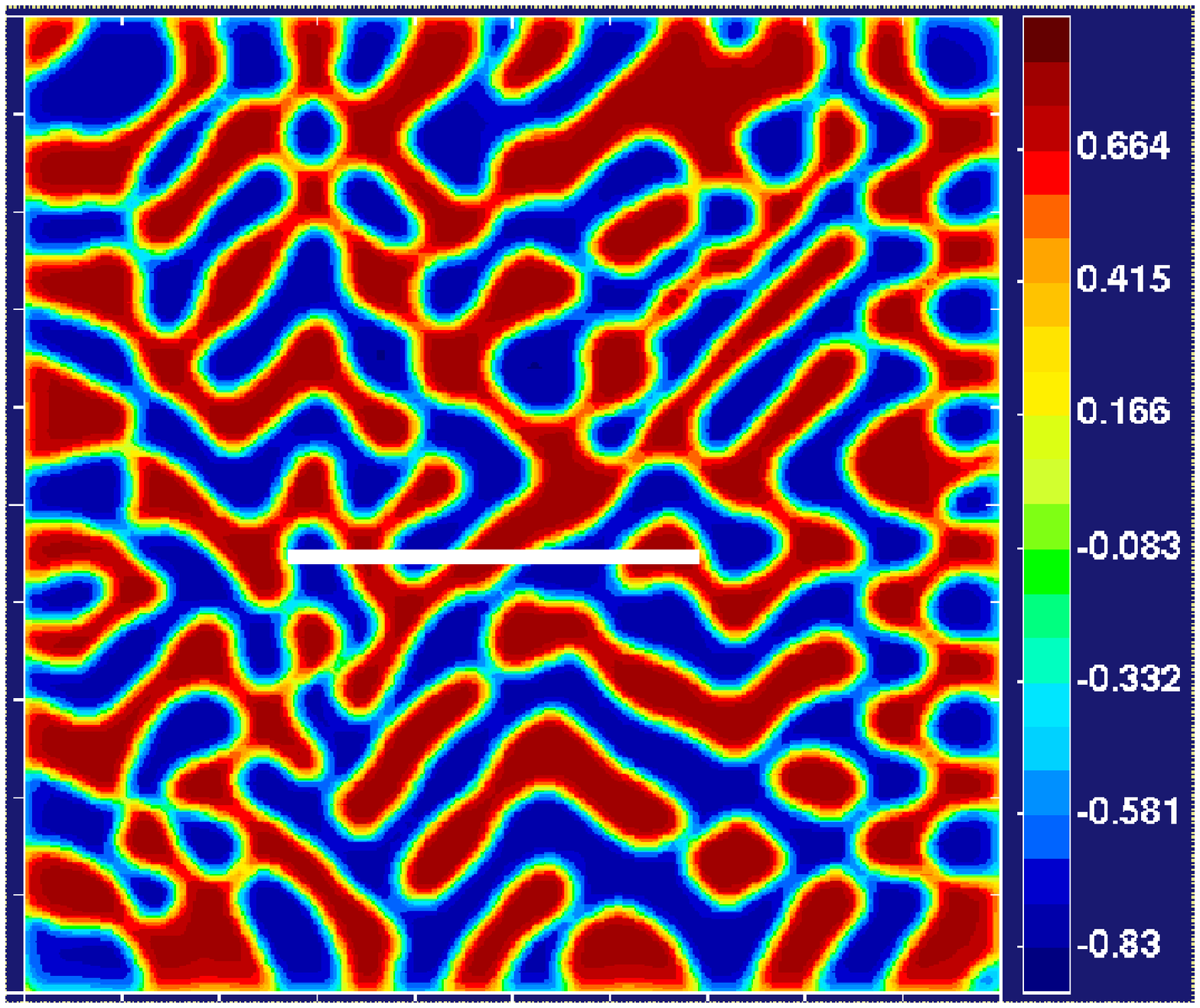}}
 \centerline{\includegraphics[width=8.0cm,angle=-0]{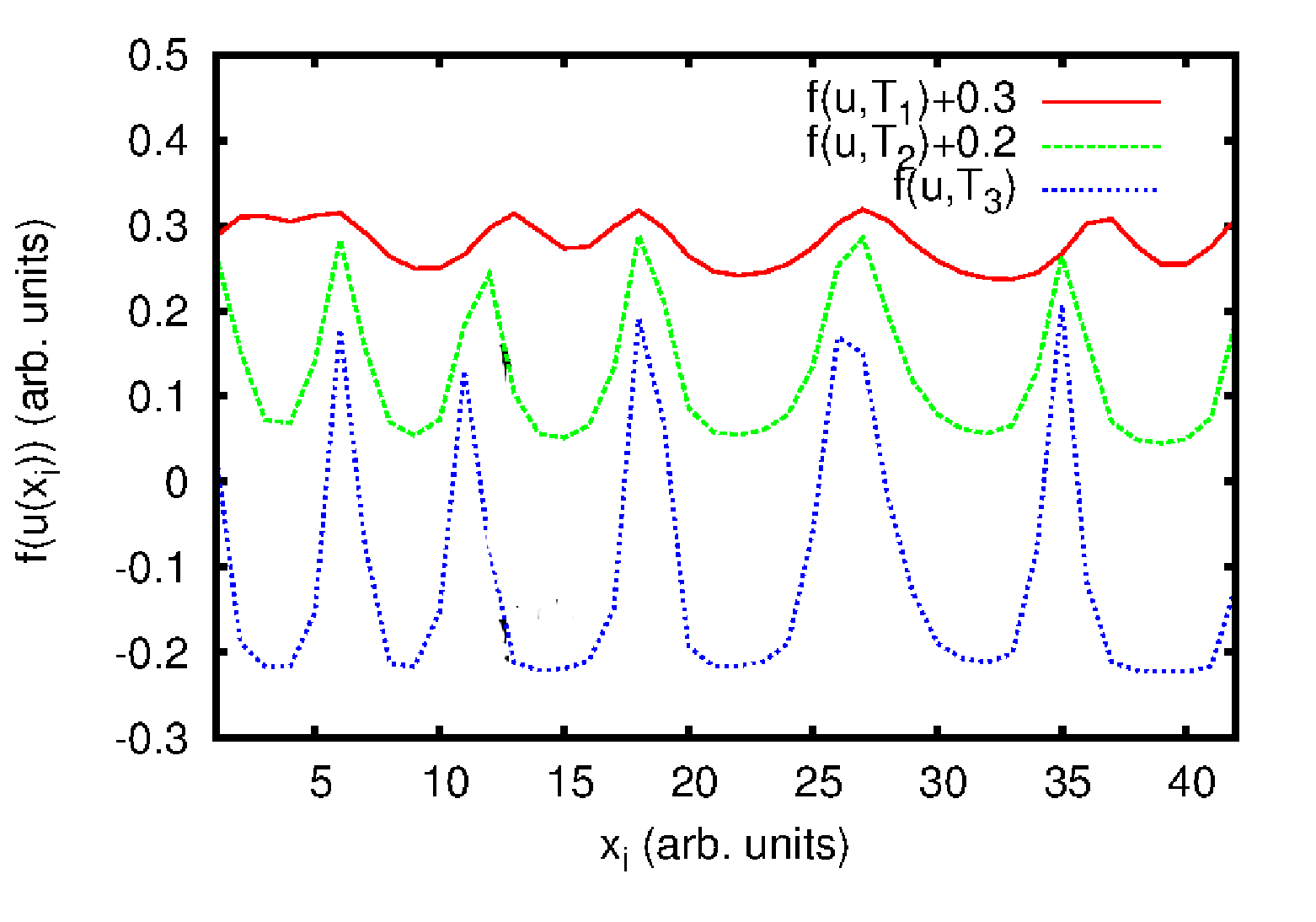}}
   \end{center}
\caption{(Color online) Top panel shows a charge density map for $p = 0.16$ constructed 
from the values of $u({\bf r}_i)$ generated from Eq.~(\ref{CH}). The red and blue colored patches 
denote regions of high and low charge density, with varying dimensions of 12 to 40~\AA.  
The lower panel shows how the spatial variation
of the GL free energy $f(u({\bf r}_i))$ (in arbitrary units) along a line of 44 sites 
(shown in white in the top panel) changes as the 
temperature is lowered from $T\lesssim T_{\rm PS}$ (top curve) to $T\gtrsim 0$ (bottom curve). 
The curves at different temperatures $T_{\rm PS} > T_1 > T_2 > T_3 $ are vertically offset for visual clarity.
Note that the low and high charge density phases occur in alternating 
free energy wells.}
\label{pMapFE} 
\end{figure}

Equation~(\ref{CH}) is solved by a semi-implicit in time finite difference scheme on a $n \! \times \! n$ square 
lattice using free boundary conditions and an initial state with small
oscillations about $u(\vec r_i)=0$.\cite{Otton} After 
examining preliminary results for  $n \! = \! 50$ to 
$n \! = \! 500$,  we settle on  $n \! = \! 100$
because this value leads to a rapid convergence 
and does not present significant finite size effects. 
The details of this method applied in one or two 
dimensions are discussed elsewhere.\cite{Otton,Mello04}

 Before considering the response to an applied field, we examine whether
the local superconducting calculations for the
simulated electronic phase separation in Sec.~II 
reproduce the well-documented low temperature gap structure measured on 
BSCCO by STM.\cite{Gomes,McElroy,Pasupathy,Pushp} These and the $T_c(p)$ 
calculations provide a framework from which the diamagnetic response 
of the normal phase can be determined.
 Usually in typical CH simulations, the value 
$A/B=1$ is chosen because it yields the maximum difference  $u=\pm \: 1$ between
the order parameters of the two phases.\cite{Otton,Mello04} Here we assume 
$A/B=|T_{\rm PS}-T|/T_{\rm PS}$,
which is approximately equal to 1 at low temperatures, and decreases towards zero 
as $T \rightarrow T_{\rm PS}$, where the charge density becomes spatially homogeneous. 
In the simulations we let the system converge to the $T \approx 0$  limit  (which corresponds to the electronic charge density 
map shown in Fig.~\ref{pMapFE}
and the histogram in the inset of Fig.~\ref{DiHist}), and subsequently lowered the ratio $A/B$ 
and let the system converge again to obtain the high
temperature behavior. Normally $0<\varepsilon \ll 1$ improves convergence
in the finite difference scheme used to solve the CH equation.\cite{Otton}
Here we used $\varepsilon=0.008$ as in previous works.\cite{EPL12a}
The exact value of the phase separation $T_{\rm PS}$ in unknown, but it is
presumably related to some of the anomalies observed in experiments probing
the normal sate.\cite{TS}  Here we assume $T_{\rm PS}=250$~K for $p=0.16$, 
since this is slightly above the onset of the temperature-dependent 
contribution to the relaxation rate of the TF-$\mu$SR signal of BSCCO
for a field of $H=7~$T applied parallel to the c-axis.\cite{Mahyari}  

As time progresses, the charge segregates into 
regions of low and high density, where the GL free energy is minimized. At the boundaries 
between the patches, the GL free energy is maximal. Figure~\ref{pMapFE} 
shows a charge density map of the two spatially separated phases of low and high density.  
Both phases have the same free energy minimum.
To provide a visual understanding of how the GL free energy changes during the phase-separation transition, 
the lower panel of Fig.~\ref{pMapFE} shows the effect of the temperature 
on $f(u)$ along a 
line of 44 sites, taken at random and shown by a white line in the top panel. 
We define $V_{gb}(T)$ to be the average height of the free energy barriers between neighboring patches.
As the temperature is lowered below $T_{\rm PS}$, the ratio of A/B increases from zero and 
the free-energy barriers grow in height, which
increases the probability of the charges being confined to specific regions.
 In general, the inter-facial free energy between the two phases has
the following temperature dependence, 
$V_{gb}(T) \! \propto \! [1-(T/T_{\rm PS})^{3/2}]$.\cite{CH}
These series of free energy wells of average depth $V_{gb}(T)$ are used in the next section
to calculate the local $d$-wave pairing amplitude.

Before finishing this section it is worth noting that the
neighboring low and high charge density 
domains displayed in Fig.~\ref{pMapFE} occur typically within 12 to 14 
lattice constants. This is less than the maximum in-plane CO correlation
length of $\xi_{\rm CO} \approx 20$ lattice constants in 
YBCO,\cite{Chang,Ghiringhelli} but larger
than $\xi_{\rm CO} \approx $ 4 to 6 lattice constants measured in 
Bi$_2$Sr$_{2-x}$La$_x$CuO$_{6+\delta}$.\cite{Comin}

\section{Calculations of Local Pairing}

After simulating a two-dimensional charge segregated state,
we use this as input for self-consistent
calculations of local Cooper pairing. The disordered state presented in Sec.~II is characterized by
a spatial variation of the free energy with many narrow wells, which can
trap charges with low kinetic energy to form {\it single particle bound states}. 
We do not have a reliable method to calculate these bound states, because they depend on the 
precise size and geometry of the patches, but on average the energy value of a bound state 
should scale with $V_{gb}(p,T)$. 

A second consequence of the above picture 
is that  {\it the clustering process provides a virtual
hole-hole attraction}. As the temperature decreases below $T_{\rm PS}$ the charges tend
to segregate, forming regions of high and low density.
In the low density regime, if two neighboring regions contains one hole each, the holes will 
``attract'' one another forming a more stable configuration of one empty region and another with two holes,
and form the patches at low temperatures shown in Fig.~\ref{pMapFE}. 
This process is similar to the exchange interaction that produces
magnetic order in that it
leads to a configuration with a lower free energy.
We can repeat this assertion with $N \! = \! 3, 4$ or 5 holes to show that $N$-body ``virtual interactions'' 
may be set in the clustering process. The $N$-body interactions are possibly very weak for $N \! > \! 2$ 
and hence apparently have no physical consequences. Of particular importance is the two-body interaction, 
because this leads to Cooper pair formation and the possibility of local superconductivity.

In the presence of such a two-body interaction, we have calculated the local $d$-wave pair potential 
$\Delta_d({\bf r}_i)$ for the spatially inhomogeneous charge density 
generated in Sec.~II by
the self-consistent Bogoliubov-deGennes (BdG) method.
This is accomplished by varying 
self-consistently the local chemical potential.  This approach somewhat resembles
that of Ghosal {\it et al,}\cite{Ghosal} who introduced a local random impurity potential
that modifies the local chemical potential. With a random distribution of chemical potential they 
obtained a random disordered charge distribution and local ($s$-wave) pairing potential
self-consistently. Here the spatial charge profile $p({\bf r}_i)$ is calculated
(in Sec. II) at a given temperature and  
kept fixed during the self-consistent procedure used to calculate 
the local chemical potential 
and the $d$-wave pairing potential at each site ``i''.
The details and the choice of kinetic parameters of the Hubbard Hamiltonian,
derived from angle-resolved photoemission spectroscopy (ARPES) experiments, are discussed in previous 
publications.\cite{PhysC12,EPL12a} The Coulomb repulsion $U$ does not influence the values of $\Delta_d$.
Here the values of $V_{gb}(p,T\approx 0)$ are chosen to yield results compatible with variations of the local energy
gap in the electronic density of states measured by STM,\cite{Gomes,McElroy,Pasupathy,Pushp,Kato,Kato2}
and the low temperature average values of the SC energy gap associated
with $T_c$ from many different experiments, as presented by H\"ufner 
{\it et al}.\cite{Hufner}

It was demonstrated before in similar calculations\cite{PhysC12,EPL12a} that 
regions with different charge densities have different values of $\Delta_d$.
 In particular, BdG calculations for low (high) charge regions 
yield lower (higher) pair amplitudes.
These calculations take much more time to converge than for a uniform system, and hence were performed over a smaller 
spatial region than the CH simulations of Sec.~II. 

\begin{figure}[ht]
     \centerline{\includegraphics[width=9.0cm,angle=0]{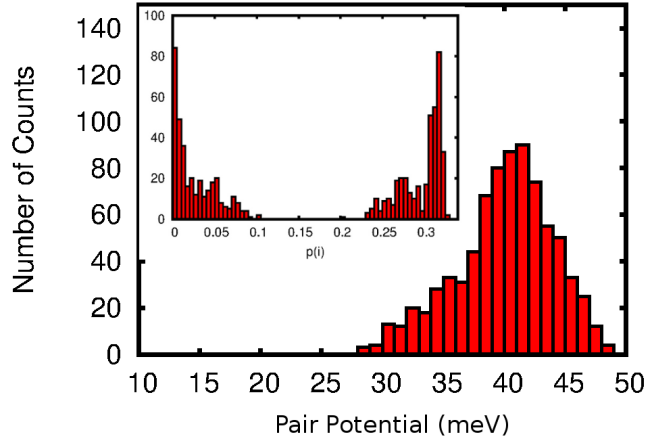}}
\caption{(Color online) Typical probability distribution for the local amplitude of
the $d$-wave pair potential $\Delta_d({\bf r}_i)$ at low temperature calculated 
by the BdG method for a square grid of $28 \times 28$ sites taken at random 
in the center of 
the charge-density map shown in Fig.~\ref{pMapFE}.
The inset shows the probability distribution of the local charge concentration. Both plots
refer to $p = 0.16$.}
\label{DiHist} 
\end{figure}

\begin{figure}[ht]
     \centerline{\includegraphics[width=9.0cm,angle=0]{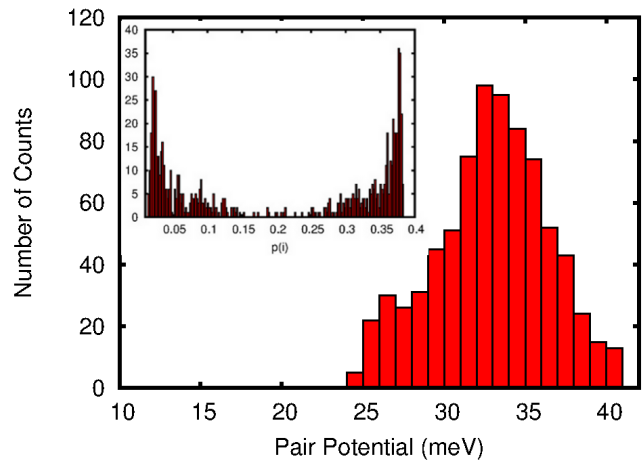}}
\caption{(Color online) The probability distribution of
the local $d$-wave pair potential $\Delta_d({\bf r}_i)$ at low temperature for 
$p = 0.20$.
The inset shows the probability distribution of the local charge concentration.}
\label{DiHistp20} 
\end{figure}
Figure~\ref{DiHist} shows the probability distribution for the local $d$-wave pairing 
amplitude $\Delta_d({\bf r}_i)$ stemming from the BdG calculations for the spatial
variation of charge-density shown in Fig.~\ref{pMapFE}. The charge density exhibits a bimodal
distribution, as shown in the inset of Fig.~\ref{DiHist}. The probability distribution
for $\Delta_d({\bf r}_i)$ resembles the gap distributions observed in underdoped to overdoped samples of
BSCCO above and below $T_c$,\cite{Gomes,Lang} but is skewed to the left
rather than to the right. The values of the local $d$-wave pairing 
amplitude calculated from the BdG mean-field approach tend to be bunched up closely and symmetrically around the 
average value over the entire system
 $\Delta^{\rm Av}_d$. For the situation here, where half the regions have low values of charge density 
distributed near $p = 0$, and half have values more closely distributed
near $p = 0.32$, the $d$-wave pairing amplitude distribution should be symmetric around
$\Delta_d \! \approx \! 40$~meV, which is the calculated value when it is assumed that $p \! = \! 0.16$
everywhere. However, the calculation underestimates the value of $\Delta_d$ for charge densities near $p \! = \! 0$,
resulting in a wider distribution of values below $\Delta^{\rm Av}_d \! \approx \! 40$~meV. Hence
the negative skewness of the distribution in Fig.~\ref{DiHist} is an artifact of the calculation.
Here we emphasize that our model makes no assumptions about the physical origin of the charge inhomogeneity, and assumes
mirror distributions of charge density above $p \! = \! 0$ and below $p \! = \! 0.32$.
On the other hand, the positively skewed gap distributions observed in BSCCO indicates that the source of the inhomogeneity
in the real material manifests itself as a non-symmetric distribution of charge density.   
 We also present similar calculations for $p \! = \! 0.20$ in Fig.~\ref{DiHistp20}. 
The insets of Fig.~\ref{DiHist} and ~\ref{DiHistp20} show
that the two distinct peaks in the probability distribution of the
charge are more pronounced for $p = 0.16$ than for $p = 0.20$, since $T_{PS}(0.20) < T_{PS}(0.16)$. The  resulting
local pair potential distribution shown in Fig.~\ref{DiHistp20}  
is also in agreement with the 
gap distribution observed by STM.\cite{Gomes}

\begin{figure}[ht!]
\begin{center}
     \centerline{\includegraphics[width=5.0cm,angle=-90]{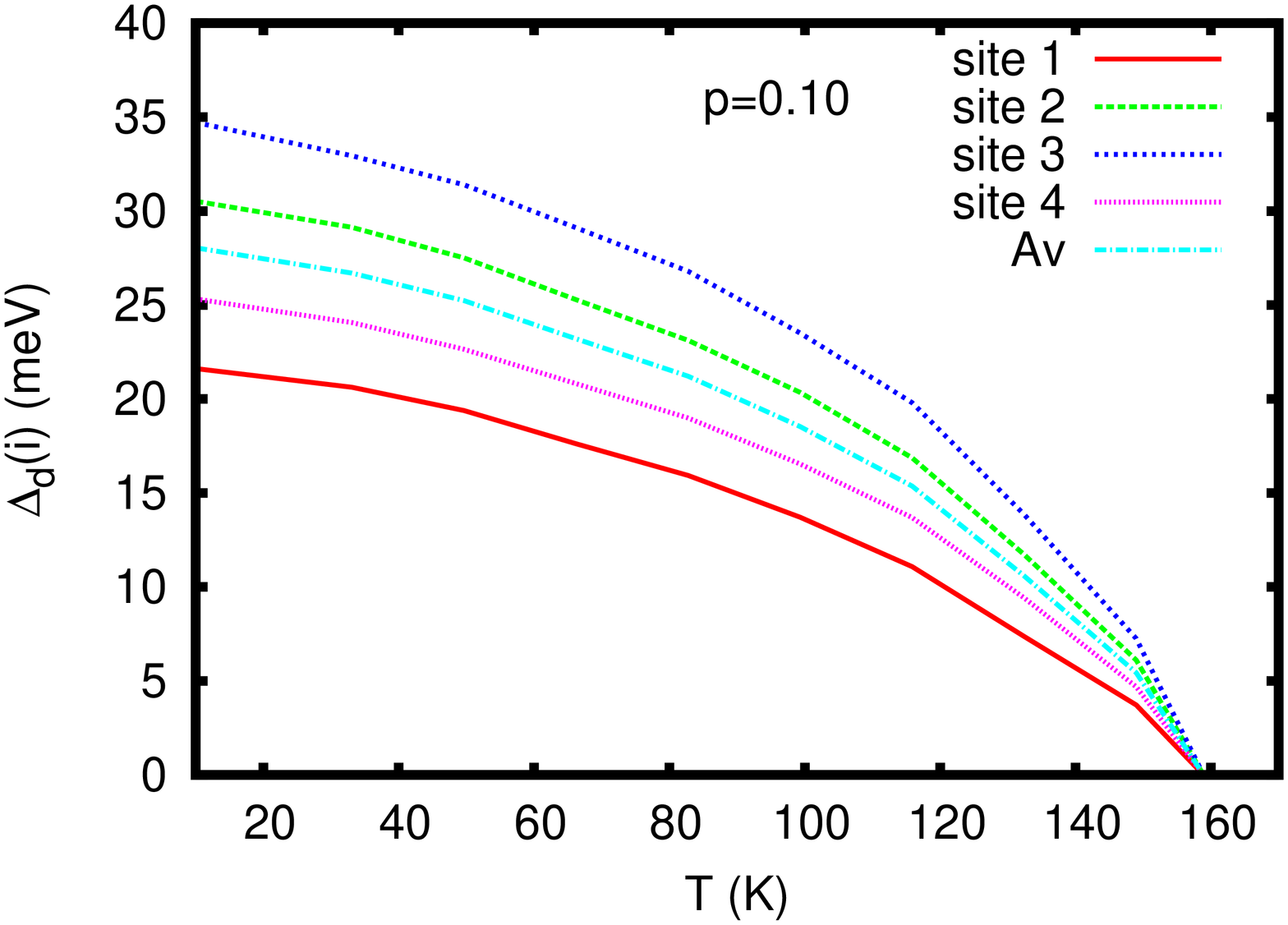}}
\vspace{0.45cm}
    \centerline{\includegraphics[width=5.0cm,angle=-90]{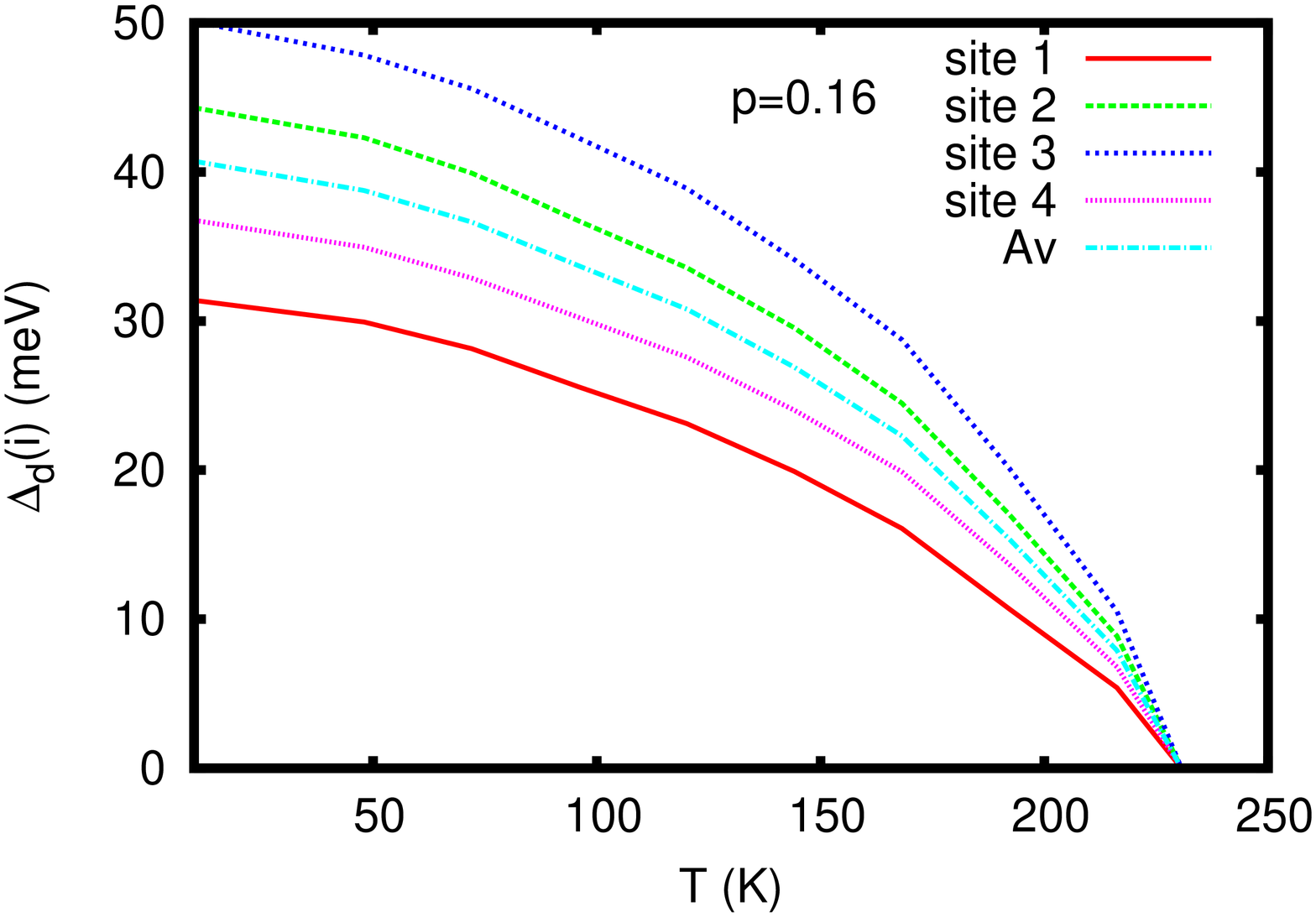}}
    \vspace{0.45cm}
    \centerline{\includegraphics[width=5.0cm,angle=-90]{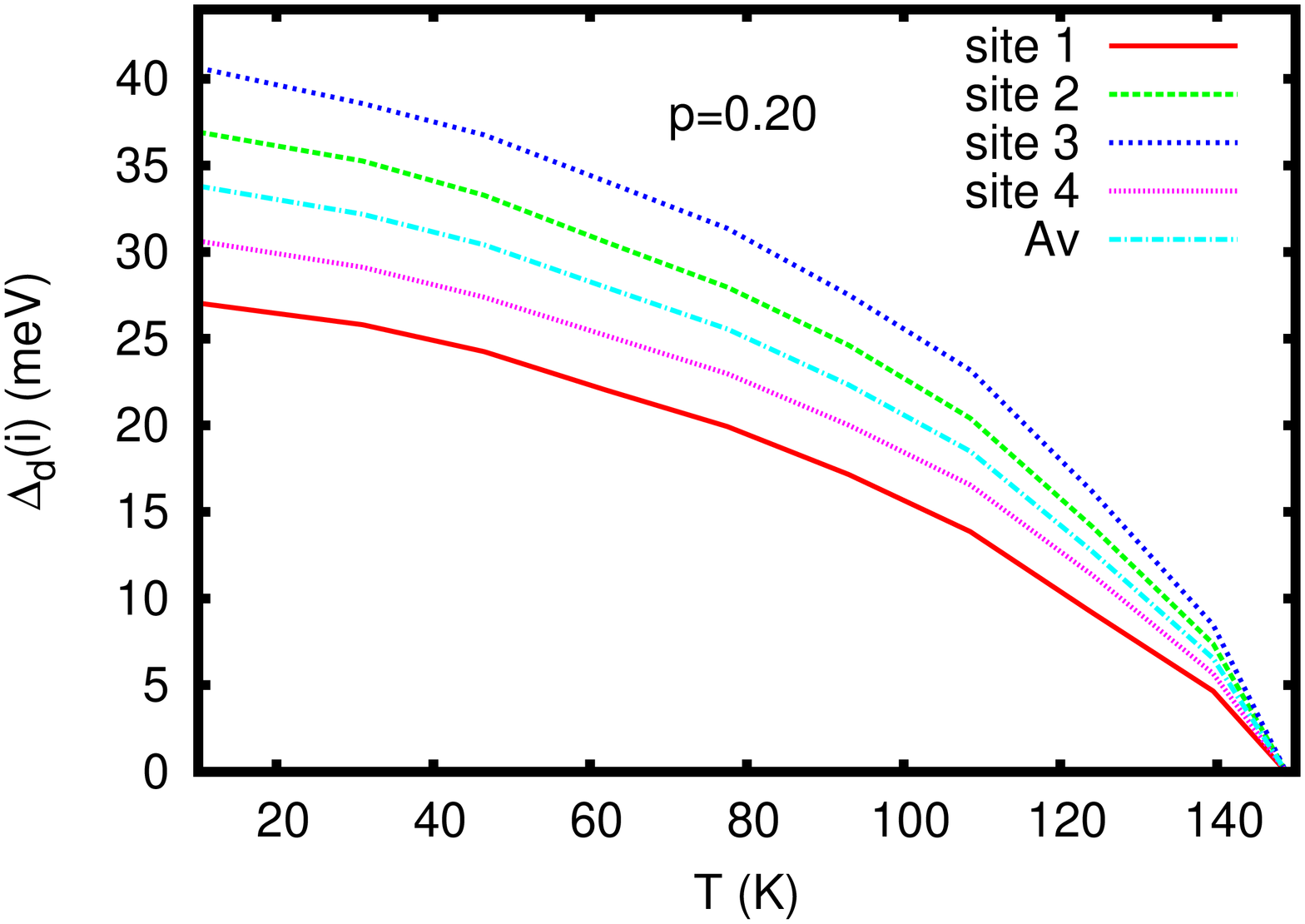}}
   \end{center}
\caption{(Color online) Temperature evolution of the $d$-wave pair potential
$\Delta_d$  at four selected locations 
(two high and two low charge density regions) together with
the average values over the entire system 
$\Delta_d^{\rm Av}$, for three different  average
dopings, $p \! = \! 0.10$, 0.16 and 0.20. The results, together with
those of Fig.~\ref{DiHist}, partially simulate the
distribution of pairing gaps in BSCCO detected by STM.\cite{McElroy,Gomes,Pushp}
The doping dependence comes from matching the $T=0$~K values 
of  $\Delta_d^{\rm Av}$ to the energy gap values
given in Refs.~\onlinecite{Gomes} and \onlinecite{Hufner}.}
\label{Deltas} 
\end{figure} 

To show representative results of the calculations outlined above, 
we plot in Fig.~\ref{Deltas} the temperature dependence of
the $d$-wave pair potential at four random locations, along with the 
average value of $\Delta_d$ for three different hole-doping 
concentrations ($p \! = \! 0.10$, 0.16 and 0.20).
The calculations are intended to simulate the situation for BSCCO measured
by STM.\cite{McElroy,Gomes,Pushp} The value of $V_{gb}(p,T\sim 0)$ determines
the average of the pairing gaps and the spread shown in Fig.~\ref{DiHist}
is naturally obtained from the simulated variations of the local charge densities.  
The zero temperature extrapolated average $\Delta_d^{\rm Av}(p,T\rightarrow 0)$
displayed in Fig.~\ref{Deltas} for these three dopings are approximately the values
of the pairing energy gap determined by Gomes {\it et al}\:~\cite{Gomes} and
estimated from various different experiments.~\cite{Hufner} 
It is important to point out that the pairing gap in 
underdoped compounds only manifests itself as a subtle gap 
feature in the STM spectra contained within the wider and more recognizable
pseudogap. Consequently, there is a greater degree of uncertainty
in the estimates of $\Delta_d^{\rm Av}(p)$ for underdoped samples.

The temperature evolution  of pairing gaps at distinct 
positions near the surface of BSCCO is such that different sized 
gaps vanish at different temperatures 
above $T_c$.\cite{Pasupathy,Gomes} 
The behavior is only partially captured
by the calculated values of $\Delta_d({\bf r}_i)$ shown 
in Fig.~\ref{Deltas}. The BdG calculations on an electronically
disordered system yield pair potentials of
varying magnitude in agreement with the experiments, 
but they all vanish at the same temperature 
due to the mean field approximation.

\section{Independent Patch Model}
    
 STM\cite{Pasupathy,Gomes} and ARPES\cite{Damascelli,Kanigel} experiments 
indicate that all $\Delta_d({\bf r}_i)$ vanish at different temperatures,
some just above and some notably higher than $T_c$.                            
Regions of local pairing may sustain shielding  supercurrents
that result in a spatially inhomogeneous diamagnetic response 
to an applied field. This is presumably the source of the  
inhomogeneous magnetic response detected in cuprates above $T_c$ 
by TF-$\mu$SR.\cite{Jeff} 

The average penetration of magnetic field into the SC patches can be simulated by 
a generalization of the critical-state Bean model.~\cite{Ketterson} 
The original work of Bean assumed a uniform superconductor in which the current density in the regions through which 
the magnetic field
penetrates is equal to its critical value $J_c$, with the simplified assumptions that $J_c$ is a constant and 
independent of magnetic field.
The critical current density is dependent on the superfluid density $n_s$ and the SC energy gap $\Delta$ via 
the following relation\cite{Schmidt}
\begin{equation}
J_c = n_s e v_c \approx n_s e \frac{\Delta}{m_e v_F} \, ,
\end{equation}
where $v_c$ is the maximum velocity of a Cooper pair, and $v_F$ is the electron velocity on the Fermi surface. 
Applying this approximation here, the critical current density associated with the SC patches is such that
$J_c({\bf r}_i) \! \propto \! n_s({\bf r}_i) \Delta_d({\bf r}_i)$.  The critical temperatures of the
SC patches exceed the bulk superconducting value $T_c$, and in the underdoped regime 
we expect $T_c({\bf r}_i) \! \propto \! n_s({\bf r}_i)$.\cite{Uemura}
From ARPES measurements on BSCCO, Anzai {\it et al}, have
determined the following relationship\cite{Anzai} 
\begin{equation}
2\Delta_{\rm N} = 8.5k_{\rm B}T_c  \propto \Delta^*\sqrt{n_s}, 
\label{relDTc}
\end{equation}
where $\Delta_{\rm N}$ is the nodal energy gap that tracks $T_c$,
and $\Delta^*$ is the gap
near the antinodes that tracks the pseudogap temperature $T^*$. 
Hence we assume $n_s({\bf r}_i) \propto [\Delta_d({\bf r}_i)/T^*(p)]^2$,
which leads to
\begin{equation}
J_c({\bf r}_i) \propto n_s({\bf r}_i) \Delta_d({\bf r}_i) \propto 
\Delta_d({\bf r}_i)^3/T^*(p)^2.
\label{Jc}
\end{equation}

We note that $2\Delta_{\rm N} = 8.5k_{\rm B}T_c$ is close to 
$2\Delta_{\rm p} = 7.9k_{\rm B}T_{\rm p}$ determined by STM,\cite{Gomes} which relates 
the value of the pairing gap to its onset temperature $T_{\rm p}$,
further justifying the use of Eq.~(\ref{relDTc}).
By applying the following Maxwell 
equation~\cite{Ketterson} 
\begin{eqnarray} 
\nabla \times {\bf B({\bf r}}_i) = \frac{4\pi}{c} {\bf J}_c({\bf r}_i,T) ,
\label{BMax} 
\end{eqnarray} 
we next 
simulate the temperature-dependent inhomogeneous line broadening observed
in optimally-doped BSCCO above $T_c$ by TF-$\mu$SR, for a static magnetic field 
applied perpendicular to the CuO$_2$ planes 
(defined here as the $z$-direction).\cite{Mahyari}

\begin{figure}[ht]
      \centerline{\includegraphics[width=9.0cm,angle=-0]{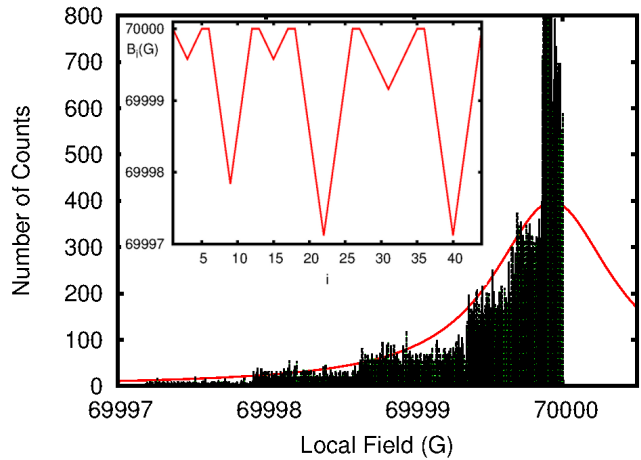}}
\caption{(Color online) (a) Calculated probability distribution of the 
internal magnetic field $n(B)$ of optimally-doped BSCCO at $T \! = \! 100$~K and
$H \! = \! 7$~T assuming a HWHM of 0.7G as determined experimentally.\cite{Mahyari} 
The red curve is a fit to a Lorentzian distribution of fields. 
The inset of (a) shows the spatial variation of $B$ over 44  of the sites used in the simulation
of the field distribution, as explained in the main text.} 
\label{BeanBHistGap} 
\end{figure}

\begin{figure}[ht]
      \centerline{\includegraphics[width=6.0cm,angle=-90]{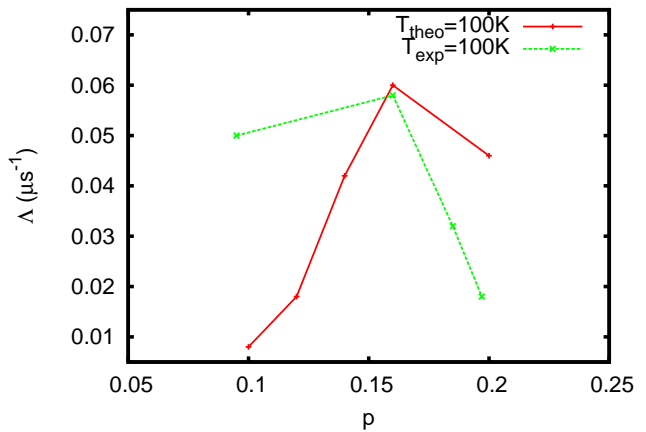}}
\caption{(Color online)  Doping dependence of $\Lambda$ at $T=100$K 
calculated by  Eqs.~(\ref{Jc}) and (\ref{BMax}) as described in the main text.
The calculations are compared to the 
experimental values of $\Lambda$ for BSCCO from Ref.~\onlinecite{Mahyari}.
}
\label{LambdaD3T2} 
\end{figure}

Equations (\ref{Jc}) and (\ref{BMax}) provide a way to estimate the field derivatives
for a given system. Then we use the Bean model and assume
a linear screening of the applied field in each patch. The absolute
response to the field depends on an unknown constant in Eq.~(\ref{Jc}), which  
can be determined by comparing the probability field histogram with
that of the experimental magnetic field distribution.
Sampling the local solutions of $\Delta_d({\bf r}_i)$, like those plotted in Fig.~\ref{Deltas}, 
we find the largest $\Delta_d({\bf r}_i)$ for the high and low density phases,
since these yield the largest values of $J_c({\bf r}_i)$ and the field derivatives.
Then we use these values to simulate the field in the 
six patches, three with high and three with low density, along the white line of Fig.~\ref{pMapFE}
(Note that the size of the patches in the inset of Fig.~\ref{BeanBHistGap}(a) are different,
but all high (low) charge density regions have the same field derivative).
Multiplying these calculations by 2000 random numbers less than one we simulate 
the results of 12,000 patches and generate a probability field histogram that can be compared 
with the measured HWHM of the Lorentzian field distribution that corresponds
to the exponential TF-$\mu$SR relaxation rate $\Lambda$.\cite{Mahyari}
For $T=100~$K and $p=0.16$ we obtain a histogram with a HWHM that corresponds to 
$\Lambda \approx 0.06\mu$s$^{-1}$ as shown in Fig.~\ref{BeanBHistGap}. From this value
and Eqs.~(\ref{Jc}) and (\ref{BMax}) we can obtain $\Lambda$ as function of the doping $p$.
The results shown in Fig.~\ref{LambdaD3T2} do not accurately
reproduce the experimental
results for BSCCO, 
especially in the underdoped regime.

Our method described above is clearly only an approximate
scheme to obtain the measured values of $\Lambda$.
As the STM experiments show,\cite{Gomes,Pasupathy} some regions
become gapless at temperatures far above $T_c$, which is not
captured by our mean field calculations.
Furthermore, an assumption implicit in Eq.~(\ref{Jc}) is that the SC patches
are completely isolated. In the next section we remove this constraint.

\section{Coupled Patches Model}

The lower panel of Fig.~\ref{pMapFE}
shows that the free energy develops barriers between the patches during the 
phase separation process, segregating the charge into distinct regions.
These regions
form independent small SC domains, which may interact 
with one another via Josephson coupling,\cite{PhysC12}
as occurs in a granular superconductor.\cite{Ketterson} 

The Josephson coupling energy is proportional to 
the maximum supercurrent that can flow between the patches ({\it i.e.}
the Josephson critical current $I_{\rm cr}$ between two SC patches) 
such that $E_{\rm J} \! = \! \hbar I_{\rm cr}/2 e$ ---
as studied in theoretical detail for a weak link between $d$-wave superconductors.\cite{Barash,Bruder} 
In particular, Bruder {\it et al.}\cite{Bruder} calculated the 
supercurrent tunneling matrix elements from second-order perturbation theory for two $d$-wave superconductors (1 and 2) of
different orientation. Clearly
the critical current $I_{\rm cr}$ is dependent on the relative orientation of the 
SC order parameter.

Here we are interested in electronic patches of $d$-wave pairing in the CuO$_2$ planes of single crystals, and
hence the special case of Josephson coupling between regions where the  
$d$-wave order parameter has the same orientation. This is different
than a true granular superconductor comprised of randomly oriented microscopic
superconducting grains.\cite{Lang}
Then
the amplitude of the $d$-wave pair potential on either side of the Josephson junction is    
\begin{equation}
 \Delta_{d,1/2}(T,\phi)=\Delta(T)\cos(2\phi) \, ,
\label{Dangle}
\end{equation}
where $\phi$ is the azimuthal angle in the $a$-$b$ plane. Equation~(\ref{Dangle}) shows that the dominant 
contribution to the critical
current is from tunneling across the non-superconducting barrier between the lobes 
of the  $d$-wave gap function on either side, and like a $s$-wave superconductor depends
mainly on $\Delta(T)$. The temperature dependence of $I_{\rm cr}$ 
for the case of interest here ({\it i.e.} two equally oriented coupled 
$d$-wave SC patches)
qualitatively resembles that of an $s$-wave superconductor.\cite{Bruder} 
Consequently, as a first approximation we assume the average Josephson coupling energy $E_{\rm J}^{\rm Av}$ to
be the simple analytical expression derived for coupling between two similar $s$-wave superconductors\:\cite{AB}
\begin{eqnarray} 
E_{\rm J}^{\rm Av}(p,T) = \frac{\pi h\Delta^{\rm Av}_d(p,T)}{2 e^2 R_n(p)}
\tanh \left[\frac{\Delta^{\rm Av}_d(p,T)}{2k_{\rm B}T} \right] \, ,
\label{EJ} 
\end{eqnarray}
where $\Delta^{\rm Av}_d(p,T) \! = \! \sum_i^N\Delta_d({\bf r}_i,p,T)/N$ is the average energy gap of all 
regions of Cooper pairing (plotted in Fig.~\ref{Deltas}), and $R_n(p)$ is the average normal resistance between 
neighboring patches at a temperature just above the phase coherence temperature
$T_c(p)$. 
It is reasonable to assume $R_n(p)$ is proportional to the normal state in-plane resistivity $\rho_{ab}(p,T\gtrsim T_c)$ just above $T_c$.
The measured in-plane resistivity $\rho_{ab}(p,T)$ for LSCO~\cite{Takagi,Ando} and BSCCO\cite{Watanabe} have 
similar values and exhibit a continuous variation over a wide range of doping.
On the other hand, the in-plane resistivity of YBCO is strongly anisotropic, with a reduced value in the
$b$-axis direction due to conductivity in the CuO chains.\cite{Ando,Ando1} Despite this difference, the $a$-axis resistivity 
$\rho_a(p,T\gtrsim T_c)$ for YBCO has the same order of 
magnitude as $\rho_{ab}(p,T\gtrsim T_c)$ for LSCO and BSCCO.

\begin{figure}[ht]
\begin{center}

\centerline{\includegraphics[width=6.0cm,angle=-90]{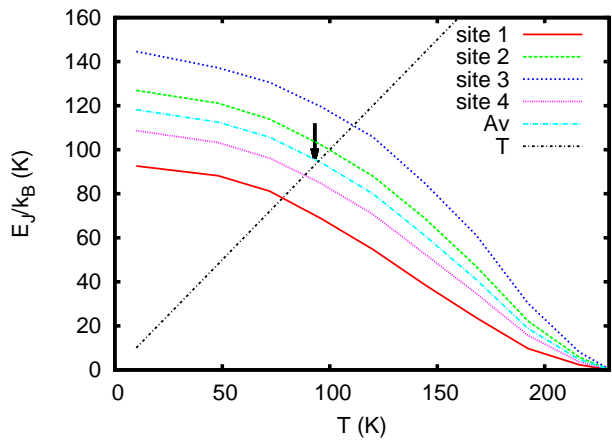}}
   \end{center}
\caption{(Color online) The  local Josephson coupling energy $E_J({\bf r}_i,p=0.16,T)$ and its 
average value for four different
sites at locations ${\bf r}_i$, calculated using Eq.~(\ref{EJ}) and the temperature dependence of the 
$d$-wave pair potential $\Delta_d({\bf r}_i)$ in Fig.~\ref{Deltas}. The arrows indicate $T \! = \! T_c$, 
corresponding to the condition $E_{\rm J}^{\rm Av}(T) \! = \! k_{\rm B} T$ and the onset of global phase coherence. 
The results are from calculations assuming a gap distribution appropriate for BSCCO, as in Sec. III.}
\label{Ejs} 
\end{figure}

\begin{figure}[ht]
\begin{center}
     \centerline{\includegraphics[width=6.0cm,angle=-90]{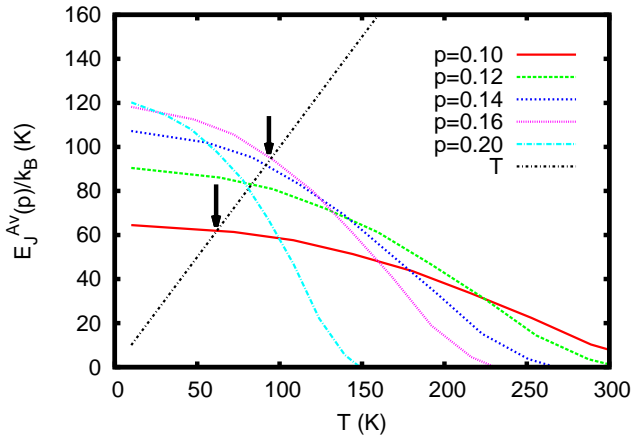}}
   \end{center}
\caption{(Color online) Temperature dependence of the average Josephson coupling energy $E_{\rm J}^{\rm Av}(p,T)$ 
for $p=0.10$, 0.12, 0.14, 0.16 and 0.20, calculated using Eq.~(\ref{EJ}) and the temperature dependence of 
$\Delta_d^{\rm Av}$ shown in Fig.~\ref{Deltas}. The arrows indicate $T \! = \! T_c$ for $p=0.16$ and $p=0.10$ 
to illustrate the condition $E_{\rm J}^{\rm Av}(p, T) \! = \! k_{\rm B}T$. 
}
\label{Ejps} 
\end{figure}

As the temperature is lowered, thermal fluctuations diminish, and long-range phase coherence 
is achieved when $k_{\rm B} T \! \approx \! E_{\rm J}^{\rm Av}(T)$ at $T_c$.\cite{EPL12b,EPL13}
Using Eq.~(\ref{EJ}) we find the value 
of $R_n(p=0.16)$ that yields $E_{\rm J}^{\rm Av}/k_{\rm B} \! = \! T_c \! \approx \! 92$~K (indicated
by the arrow in Fig.~\ref{Ejs}), which is the
approximate value of $T_c$ for optimally-doped YBCO and BSCCO. Due to the inhomogeneous charge distribution
the Josephson coupling energy change at different locations and we plot also the results at different
patches in Fig.~\ref{Ejs}) that will be  used in the next section to calculate the field response.
We then obtain $T_c(p,T)$ for other dopings via the
ratio $\rho_{ab}(p)/\rho_{ab}(p=0.16) \! = \! R_n(p)/R_n(p=0.16)$ for BSCCO (or LSCO). Clearly this procedure, 
which is shown in Fig.~\ref{Ejps}, will not reproduce the 60~K plateau of $T_c(p)$ for YBCO near $p \! = \! 1/8$.

The results for 
$p \! = \! 0.10$, 0.12, 0.14 and 0.20 are displayed in Fig.~\ref{Ejps} and 
they are in agreement with the measured resistivity
transitions. 
It is important to emphasize that the $T_c(p)$ results are robust to
the details of the pairing potential in the underdoped region. Since the 
normal resistivity $R_n(p)$ rapidly increases as $p$ decreases towards
$p=0.05$, it is easy to see from Eq.(\ref{EJ}) that
the average Josephson energy $E_{\rm J}(p\approx 0.05) \rightarrow 0$. 
This is an important feature of our approach,
because as already mentioned, the values of the normal-state pairing gaps
are difficult to determine by STM for the underdoped compounds.

The assumption that $T_c(p)$ is controlled by the Josephson coupling energy
and the fact that the exponential relaxation rate $\Lambda(p,T)$  
roughly tracks $T_c(p)$,\cite{Sonier:08,Mahyari} 
imply that the local
Josephson coupling energy $E_{\rm J}({\bf r}_i,p,T)$ and the inter-patch
tunneling currents play important roles in the doping dependence of 
the magnetic response of the system. 
With this in mind we replace  ${\bf J}_{c}({\bf r}_i,T)$ in Eq.~(\ref{BMax}) with the 
Josephson critical current density ${\bf J}_{\rm cr}({\bf r}_i,T)$ as follows
\begin{eqnarray} 
\nabla \times {\bf B({\bf r}}_i) = \frac{4\pi}{c} {\bf J}_{cr}({\bf r}_i,T) \, .
\label{BMaxEj} 
\end{eqnarray} 
In this case the spatial derivatives of the field inside the patches will
be proportional to the local Josephson coupling energy $E_J({\bf r}_i,p,T)$
since $E_J=\hbar I_{\rm cr}/2e$, 
rather than the cube of the local pair potential $\Delta({\bf r}_i,p,T)$
as in Section IV.  Typical results of $E_J({\bf r}_i,p,T)$ used in the calculations
are plotted in Fig.~\ref{Ejs}.
With this change, $n(B)$ is derived in
a manner similar as before for $p=0.16$ and $T=100$~K. However the results
for other temperatures and dopings
in this new approach
depend on the ratio of the average Josephson coupling energy.

\begin{figure}[ht]
      \centerline{\includegraphics[width=9.0cm,angle=-0]{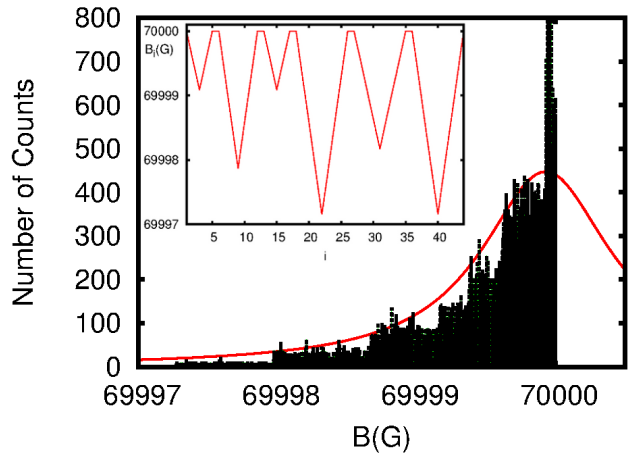}}
\caption{(Color online)
(a) Calculated probability distribution of the internal magnetic field
$n(B)$ at $T=100~$K assuming local tunneling critical currents between the patches. 
The red curve is a fit to a Lorentzian distribution of fields.
The inset  shows the spatial variation of $B$ over 44 sites.}
\label{BeanBHist} 
\end{figure}

\begin{figure}[ht]
     \centerline{\includegraphics[width=6.0cm,angle=-90]{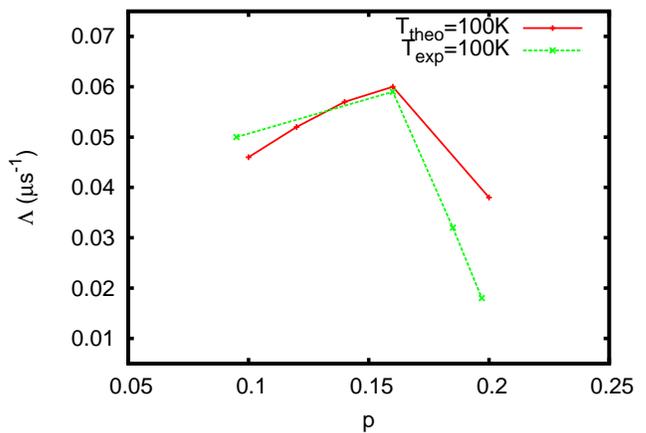}}
\caption{(Color online)  Doping dependence of $\Lambda$ at $T = 100$K
calculated by the coupled patches method described in the main text compared to the 
experimental values for BSCCO from Ref.~\onlinecite{Mahyari}.}
\label{MuSRpsTs} 
\end{figure}

We now use this model to  estimate the  diamagnetic
field response for different dopings using the 
average Josephson coupling energy curves shown in Fig.~\ref{Ejps}. 
The field derivatives of the SC patches for any doping $p$ are estimated 
from the ratio $E_{\rm J}^{\rm Av}(p=0.16,T=100$~K$)/E_{\rm J}^{\rm Av}(p,T)$. 
The resulting probability field histogram for various dopings at $T=100$~K 
and the respective values of $\Lambda(p)$ along with the experimental data 
are plotted in Fig.~\ref{MuSRpsTs}. The agreement is reasonable considering  
the approximations made in our model.

\begin{figure}[ht]
     \centerline{\includegraphics[width=6.0cm,angle=-90]{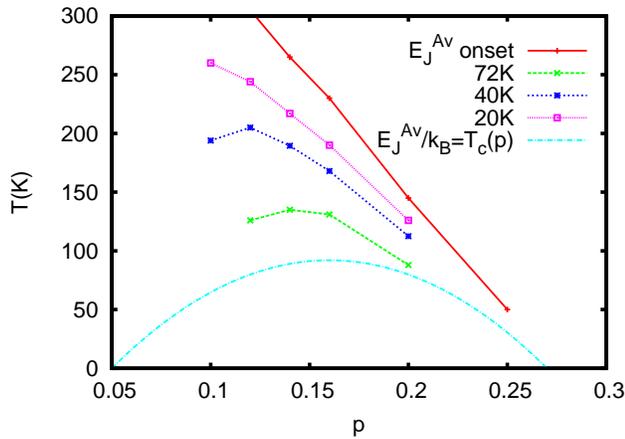}}
\caption{(Color online) Phase diagram showing curves of constant values of $E_{\rm J}^{\rm Av}/k_{\rm B}$ 
as a function of temperature and hole doping $p$. The $T_c(p)$ curve was generated from 
Fig.~\ref{Ejs}. The value at which the $d$-wave pairing
amplitude $\Delta_d(p)$ becomes nonzero corresponds to the onset curve of $E_{\rm J}$,
which is close to the pseudogap temperature $T^*(p)$.
The curve $E_{\rm J}/k_{\rm B} \! = \! 40K$ resembles the doping dependence of 
$\Lambda \! = \! 0.02~\mu$s$^{-1}$ above $T_c$
observed in the TF-$\mu$SR experiments on BSCCO of Ref.~\onlinecite{Mahyari}.}
\label{PhaseDiag} 
\end{figure}
\section{Discussion}

Now that we have shown that the Josephson coupling energy can be used to
derive the doping dependent magnetic field response measured in the
TF-$\mu$SR experiments, we plot in 
Fig.~\ref{PhaseDiag} the ratio $E_{\rm J}^{\rm Av}(p,T)/ k_{\rm B}$ as a function 
of hole doping and
temperature. The curve for $T_c(p)$ represents the temperatures that satisfy 
the long range phase coherence condition, {\it i.e.} 
$E_{\rm J}^{\rm Av}(p,T) \! = \! k_{\rm B} T$ (derived from Fig.~\ref{Ejps}). We also show a curve that
marks the onset of Josephson coupling. By Eq.~(\ref{EJ}), this also marks the onset of 
the pairing amplitude $\Delta_d(p)$, and hence should be close to 
the pseudogap temperature $T^*(p)$. The curve for
$E_{\rm J}^{\rm Av}/k_{\rm B} \! = \! 40$~K resembles the doping dependence of the measured value 
$\Lambda \!=\! 0.02~\mu$s$^{-1}$ above $T_c$ in 
Ref.~\onlinecite{Mahyari} with a maximum value below optimal doping.
On the other hand, the curve for $E_{\rm J}^{\rm Av}/k_{\rm B} \! = \! 72$~K resembles
the doping dependence of the onset temperature for diamagnetism detected in BSCCO by torque
magnetometry.\cite{Ong} These comparisons suggest that the sensitivity of different experimental
techniques to the magnetic response of such an inhomogeneous system depends on how strongly the individual SC
regions are coupled. As a local magnetic probe, the muon is sensitive to the diamagnetism
of weakly coupled and/or isolated dilute patches, whereas the diamagnetic signal detected 
at lower temperatures by torque magnetometry 
presumably comes from regions that are more strongly Josephson coupled.
It is also interesting to point out that for the doping range considered, the curve for the smaller value 
$E_{\rm J}^{\rm Av}/k_B \! = \! 20$~K in Fig.~\ref{PhaseDiag} is systematically higher,
but is qualitatively consistent with the doping dependence of the onset 
temperature for precursor pairing inferred from $c$-axis infrared response measurements on YBCO.\cite{Dubroka}   

\section{Conclusion}

We have applied a general phase separation method to simulate small-scale electronic inhomogeneity
in the CuO$_2$ planes of high-$T_c$ cuprate superconductors.  Using the BdG equations
we have calculated the spatially-varying Cooper pair potential in this environment,
and shown that inhomogeneous superconductivity may occur in this
system. With appropriate parameters we can reproduce the 
measured doping and temperature dependences of the local and average pair
potentials of BSCCO. 
Using these results we have attempted to calculate the
observed doping dependence of the inhomogeneous magnetic field
response above $T_c(p)$ detected by TF-$\mu$SR  
via two different approaches: (i) The system is considered to be a single 
superconductor with electronic inhomogeneity whose average 
$d$-wave pair potential has a dome-shaped hole-doping dependence that yields $T_c(p)$. 
In this case the main source of the inhomogeneous
diamagnetic response comes from shielding currents around 
individual patches, assumed to scale
with the cube of the local pair potential $\Delta_d({\bf r}_i,p,T)$. 
(ii) The system is considered to be like a granular superconductor
composed of individual patches with local pairing potentials that
are connected with one another via Josephson coupling of energy $E_{\rm J}({\bf r}_i,p,T)$. 
In this case $T_c(p)$ corresponds to the onset of long-range phase order between the patches,
and the source of the inhomogeneous magnetic response comes from varying critical tunneling 
currents.  This second approach yields better agreement
with the TF-$\mu$SR measurements.

We conclude by noting a significant distinction between our model and the interpretation
of the TF-$\mu$SR measurements in Ref.~\onlinecite{Mahyari}. In the latter it is assumed
that the measured temperature-dependent relaxation rate $\Lambda$ above $T_c$ is caused by
a distribution of time-averaged local magnetic fields associated with inhomogeneous SC
fluctuations, whereas our model assumes a static distribution of magnetic field. Even so,
the formation and effective screening from patches with fluctuating phase-coherent Cooper
pairs may be controlled by the same parameters, preserving the qualitative behaviors
predicted from our model.

\section{Acknowlegdments}

We would like to thank M. P. Kennett and D. G. Hawthorn for informative discussions.
EVLM acknowledges partial financial aid from the Brazilian agencies Capes, CNPq and FAPERJ.
JES acknowledges support from the Natural Sciences and Engineering Research Council of
Canada, and the Canadian Institute for Advanced Research.

\end{document}